\documentclass[twocolumn,preprintnumbers,amsmath,amssymb,superscriptaddress]{revtex4-1}

\usepackage{graphicx}
\usepackage{bm}
\usepackage{color}
\usepackage{ulem}
\usepackage{multirow}
\usepackage{lipsum}

\begin{document}

\title{Soliton refraction through an optical soliton gas}
\author{Pierre Suret}
\affiliation{Univ. Lille, CNRS, UMR 8523 - PhLAM -
  Physique des Lasers Atomes et Mol\'ecules, F-59 000 Lille, France}
\author{Martin Dufour}
\affiliation{Univ. Lille, CNRS, UMR 8523 - PhLAM -
  Physique des Lasers Atomes et Mol\'ecules, F-59 000 Lille, France}
\author{Giacomo Roberti}
\affiliation{Department of Mathematics, Physics and Electrical Engineering, Northumbria University, Newcastle upon Tyne, NE1 8ST, United Kingdom}
\author{Gennady El}
\affiliation{Department of Mathematics, Physics and Electrical Engineering, Northumbria University, Newcastle upon Tyne, NE1 8ST, United Kingdom}
\author{Fran\c{c}ois Copie}
\affiliation{Univ. Lille, CNRS, UMR 8523 - PhLAM -
  Physique des Lasers Atomes et Mol\'ecules, F-59 000 Lille, France}
\author{St\'ephane Randoux}
\email{stephane.randoux@univ-lille.fr}
\affiliation{Univ. Lille, CNRS, UMR 8523 - PhLAM -
  Physique des Lasers Atomes et Mol\'ecules, F-59 000 Lille, France}

\date{\today}

\begin{abstract}
We report an optical fiber experiment in which we investigate the interaction between an individual  soliton and a dense soliton gas. We evidence a refraction phenomenon where the tracer soliton experiences an effective velocity change due to its interaction with the optical soliton gas. This interaction results in a significant spatial shift that is measured and compared with theoretical predictions obtained via the inverse scattering transform (IST) theory. The effective velocity change associated with the refraction phenomenon is found to be in good quantitative agreement with the results of the spectral kinetic theory of soliton gas.   Our results validate the  collision rate ansatz that plays a fundamental role in the kinetic theory of soliton gas and also is at heart of generalized hydrodynamics of many-body integrable systems.
\end{abstract}


\maketitle

 Solitons are localized nonlinear wave structures that owe their existence to an exact balance between the wave's nonlinearity and the medium's dispersion. Solitons have been observed in a great variety of physical systems including  optics \cite{Hasegawa_book:95}, matter waves \cite{Nguyen:14,Luo:20}, fluids  \cite{Yuen:75,Trillo:16}, metamaterials \cite{Deng:17}, biophysics \cite{Heimburg:05}. Solitons play a fundamental role in nonlinear physics due to the remarkable property of retaining their shape, amplitude and velocity upon interactions with other solitons. 

The interaction between solitons is a complex nonlinear process that can be understood within the framework of the celebrated inverse scattering transform (IST)   developed to solve integrable nonlinear partial differential equations like the Korteweg-de Vries (KdV) equation or the one-dimensional nonlinear Schr\"odinger equation (1D-NLSE) \cite{Ablowitz:73,Zabusky:65,Novikov_book,Remoissenet_book,yang2010nonlinear}. Considered on sufficiently large spatiotemporal scales solitons  behave as quasi-particles experiencing short-range pairwise elastic interactions accompanied by well-defined phase/position shifts. The process of elastic collision between two solitons occurs without energy exchange between them and has been studied experimentally in great detail in many physical systems \cite{Ikezi:70,Nguyen:14,Aossey:92,Slunyaev:17,Mitschke:87,Andrekson:90,Aitchison:91,Shalaby:92,Parker:08,Stellmer:08}.

Recently, the interaction between an individual (tracer) soliton and a large-scale coherent nonlinear structure such as rarefaction and dispersive shock wave has been studied both theoretically and experimentally \cite{Maiden:18,  sprenger_hydrodynamic_2018, Biondini:18, Biondini:19, sande_dynamic_2021, mucalica_solitons_2022, ablowitz_soliton-mean_2022}. The trajectory of the tracer soliton within these macroscopic nonlinear structures  is directed by the structure's mean field, which results either in the trapping of the soliton inside or its  transmission (tunneling)  through the large-scale nonlinear wave.

In this Letter, we present an optical fiber experiment where we examine the interaction between a tracer soliton and  a  dense soliton gas (SG)---a large ensemble of solitons that exhibits coherence on the microscopic (``dispersive'') scale but is incoherent on the macroscopic (``hydrodynamic'') scale.  Thus, in contrast to the fully coherent configurations examined in the previous experimental work \cite{Maiden:18}, the  soliton now interacts with a {\it random} nonlinear wave. 

\begin{figure*}[!t]
  \includegraphics[width=1\textwidth]{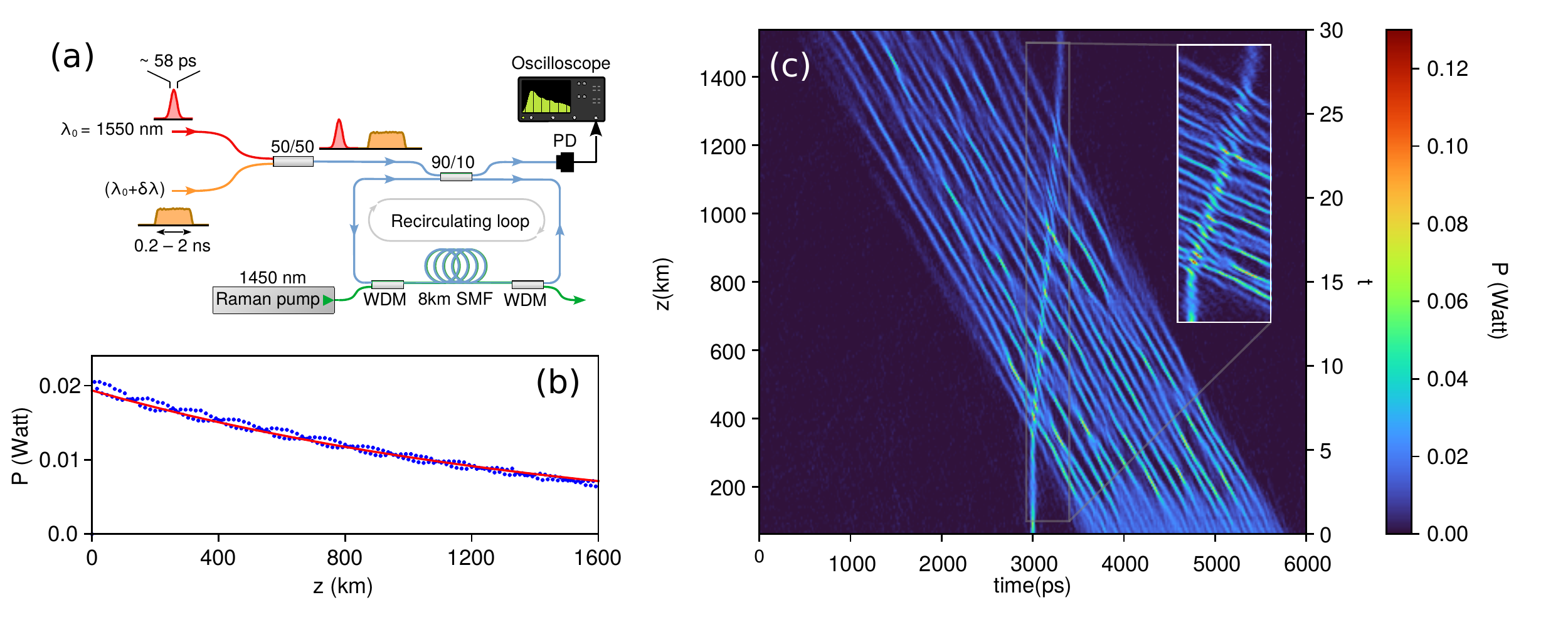}
  \vspace{-20pt}
  \caption{(a) Schematic representation of the experimental setup. (b) Measured decay (blue points) of the optical power in the recirculating fiber loop. The red line represents an exponential fit of the experimental points, giving a power decay rate of $\sim 0.0027$ dB/km. (c) Typical space-time diagram showing the refraction phenomenon due to the interaction between a tracer soliton and an optical SG forming a bound state. The tracer soliton has a duration of $\sim 58$ ps. It is shifted by $\sim 280$ ps due to its interaction with the optical SG that has a duration of $\sim 2$ ns. The secondary (right) vertical axis shows the normalized evolution time defined as $t= \gamma P_0 z /2=z/(2L_{NL})$, ($L_{NL}\sim 26.5$ km). 
}
\end{figure*}

SGs have been recently realized in optical and in water wave experiments \cite{Marcucci:19,Redor:19,Suret:20}. Nowadays, hydrodynamics and statistics of SGs is an active research area in statistical mechanics \cite{Bonnemain:22}, mathematical physics \cite{Girotti:22}, nonlinear physics \cite{Redor:21, Slunyaev:22,Pelinovsky:17}, and constitutes a new chapter of turbulence theory, termed integrable turbulence \cite{Zakharov:09}. 

In our experiment we observe that the nonlinear interaction between a tracer soliton and a finite portion of a dense SG results in the refraction phenomenon similar at a qualitative level to what is observed in ray optics at the interface between two media having different refractive indexes. Unlike the classical refraction of light rays, the observed phenomenon is  inherently nonlinear. We measure the macroscopic space shift associated with the solitonic refraction and show that it is quantitatively well described by the IST theory \cite{Alonso:85a,Alonso:85b,Borghese:18}.  Specifically, we compare the experimentally observed macroscopic position shift $\Delta x$ acquired by the tracer soliton over a large propagation distance  with the ``first-principle'' IST predictions based on the accumulation of the individual position shifts in pairwise  interactions  with solitons comprising the SG \cite{Zakharov:72}, 
\begin{equation}\label{deltax_IST}
\Delta x= \sum_{j} \Delta(\lambda_p, \lambda_j) \, .
\end{equation}
Here $\lambda_p$ is the IST spectral parameter of the tracer soliton and  $\Delta(\lambda_p, \lambda_j)$ is the position shift of  the tracer soliton due to its interaction with the soliton with the spectral parameter $\lambda_j$ within the SG.
 
Additionally, following the approach prescribed by the spectral kinetic theory of SGs \cite{GEl:05,GEl:19,GEl:21}, we measure the effective velocity of the tracer soliton 
propagating through the SG.  From the physical perspective, the effective adjustment of the soliton velocity due to the interaction with SG is described by the 
the {\it collision rate ansatz} (CRA) that extrapolates the properties of two-soliton interactions to  dense macroscopic ensembles of solitons, in which solitons never separate to exhibit individual short-range interactions. We compare quantitatively the experimental and theoretical values of the effective velocity of the tracer soliton interacting with the dense SG. This comparison represents a crucial step in the physical validation  of the CRA and ultimately, the kinetic theory of SGs as will be explained below.

\begin{figure*}[!t]
  \includegraphics[width=1\textwidth]{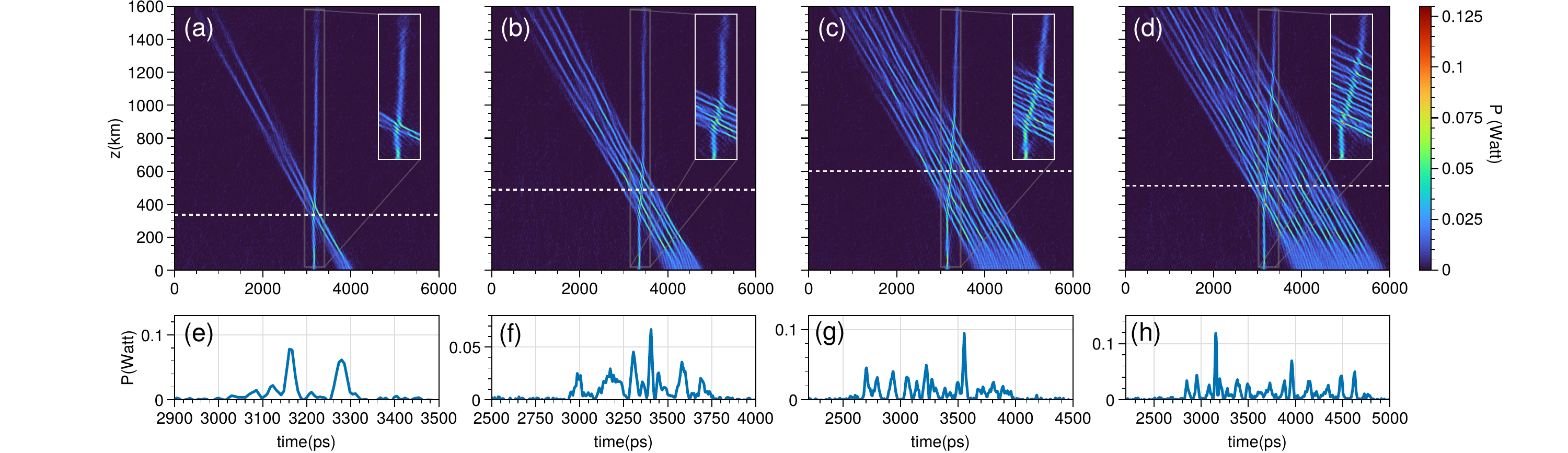}
  \vspace{-15pt}
  \caption{(a)--(d): Selected set of 4 space-time diagrams among the 29 available in one single experiment. The shift experienced by the tracer soliton grows with the initial size (duration) of the optical SGs. (e)--(h): Time evolution of the optical power at selected positions (indicated by the horizontal white dashed lines in the top row) where tracer solitons interact with the SGs. 
}
\end{figure*}

Our experimental setup is schematically shown in Fig. 1(a). It consists of a recirculating fiber loop already used in ref. \cite{Kraych:19a,Kraych:19b,Copie:22} for some experimental investigations of the nonlinear stage of modulational instability of a plane wave. The recirculating fiber loop is made up of $\sim 8$ km of single mode fiber (SMF) closed on itself by a $90/10$ fiber coupler. The coupler is arranged in such a way that $90 \%$ of the intracavity power is recirculated.  The optical signal (a tracer soliton plus a SG) circulates in the clockwise direction. At each round trip, $10 \%$ of the circulating power is extracted and directed toward a photodetector (PD) coupled to a sampling oscilloscope ($160$ Gsa/s) leading to an overall $32$ GHz detection bandwidth. Experimental data are subsequently processed numerically to construct single-shot space-time diagrams showing the wavefield dynamics. The losses accumulated over one circulation in the fiber loop are partially compensated using a counter-propagating Raman pump coupled in and out of the loop via wavelength division multiplexers (WDMs). As shown in Fig. 1(b), this reduces the effective power decay rate of the circulating field to $\alpha_{eff} \sim 6.2 \times 10^{-4}$ km$^{-1}$ or equivalently $\sim 0.0027$ dB/km. 

The optical signal propagating inside the fiber loop is composed by a pulse with a duration of $\sim 58$ ps followed by an optical SG initially in the form of a ``long'' square pulse perturbed by some optical noise, see Fig. 1(a)(c) together with Supplementary material for details about the generation of light signals.  The short pulse is initially well separated from the square pulse that evolves into a fully randomized bound state SG \cite{Gelash:19,Marcucci:19,Bonnefoy:20,GEl:16}. The short pulse and the optical SG have slightly different group velocities, resulting in collisions/interactions between the SG and pulse starting from propagation distances of $\sim 350-400$ km. In practice, the group-velocity difference $\delta v_g$ between the pulse and the optical SG is realized by using two laser fields with wavelengths that are slightly detuned by $\delta \lambda \simeq 0.125$ nm, resulting in $[\delta v_g]^{-1} \simeq -2.16$ ps/km (see Supplementary material). 

The refraction effect evidenced in Fig. 1(c) is associated with a significant shift ($\sim 280$ ps) of the position of the soliton at the spatial coordinate ($z \sim 1200$ km) from which its emerges from the SG. The inset in Fig. 1(c) clearly shows that the velocity of the tracer soliton has slightly changed after is has been transmitted through the SG. This velocity change arises from the fact that the observed dynamics is not perfectly integrable due to the presence of small dissipation in the experiment.

The optical signal circulating in the fiber loop is not only composed of one pulse and one SG but of a periodic train of $29$ short pulses, each pulse being followed by ``its own SG'', see Supplementary material showing the whole experimental pattern recorded in single shot. The short pulses are all designed to be identical but in practice, their peak power is $P_p=39.2 \pm 5.54$ mW and their duration is $\Delta T= 58 \pm 16$ ps (FWHM). The SGs have the initial form of square boxes with a mean power $P_{SG}=19.2 \pm 1.6$ mW. Their duration increases monotonically from $\sim 200$ ps to $\sim 2000$ ps in $29$ steps. Using this strategy, we capture in one single shot the space-time evolution of a set of $29$ experiments where we observe the interaction between a pulse and optical SGs with increasing extents.  

Fig. 2 shows a selection of four experiments from the set of $29$ experiments available. Whatever the initial extent of the optical SG, the solitonic refraction phenomenon is observed. The insets in Fig. 2(a)--(d) show that the time shift experienced by the soliton increases with the duration of the optical SG. Fig. 2(e)--(h) show the time signal measured at some selected propagation distances, when the tracer solitons interact with the SGs. Fig. 2(g)(h) show that the tracer soliton may reach a peak power $\sim 3$ times greater than its initial peak power ($\sim 39$ mW) due to its interaction with the SG. 

Now we compare the shift measured in the experiment with numerical simulations and the results from the IST theory. As show in ref. \cite{Kraych:19a,Kraych:19b,Copie:22}, dynamical features observed in our recirculating fiber loop can be quantitatively well described by the following 1D-NLSE with a small linear damping:
\begin{equation}\label{eq:NLSE}
  i\frac{\partial A}{\partial z}=\frac{\beta_2}{2}\frac{\partial^2
    A}{\partial T^2}-\gamma|A|^2\psi -i \frac{\alpha_{\rm
      eff}}{2} A.
\end{equation}
$A(z,T)$ represents the complex envelope of the electric field that slowly varies in physical space $z$ and time $T$.  The Kerr coefficient of the fiber is $\gamma=1.3$ W$^{-1}$km$^{-1}$. The group velocity dispersion coefficient is $\beta_2=-22$ ps$^2$ km$^{-1}$. 

\begin{figure*}[!t]
  \includegraphics[width=1\textwidth]{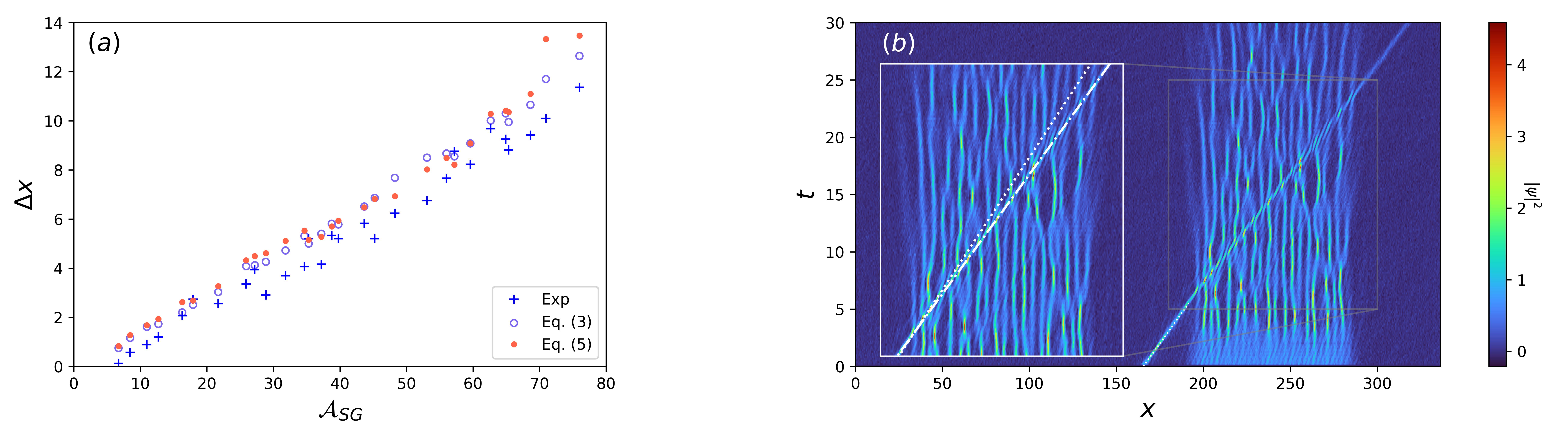}
    \vspace{-17pt}
  \caption{Comparison between experimental results and theoretical results from the IST theory and the kinetic theory of SG. (a) Shifts measured in the experiment (blue crosses), in numerical simulations of Eq. (\ref{eq:NLSE_adim}) (empty circles) and using Eq. (\ref{eq:shift}) (filled circles). (b) Space-time diagram showing the interaction between the tracer soliton and the dense SG composed of $24$ solitons in the reference frame moving at the group velocity of the SG. The dashed-dotted white line follows the trajectory of the tracer soliton inside the SG while the dotted white line follows the trajectory of tracer soliton before it interacts with the SG.
}
\end{figure*}

For a straightforward comparison with the IST theory, we use the following dimensionless form of the focusing 1D-NLSE 
\begin{equation}\label{eq:NLSE_adim}
  i \psi_t+\psi_{xx}+2|\psi|^2 \psi+i\epsilon \psi=0
\end{equation}
which is obtained from Eq. (\ref{eq:NLSE}) using the following transformations: $\psi=A/\sqrt{P_0}$, $x=T\sqrt{\gamma P_0/|\beta_2|}$, $t= \gamma P_0 z /2$, $\epsilon=\alpha_{eff}/(\gamma P_0)$. The value $P_0=29$ mW has been measured in an annex calibration experiment, see Supplemental material. Using the canonical expression of the 1D-NLSE given by Eq. (\ref{eq:NLSE_adim}), the maximum evolution time in normalized units is as large as $\sim 30$ (see secondary evolution axis in Fig. 1(c)), which corresponds to a propagation over $60$ nonlinear lengths with the nonlinear length being defined as $L_{NL}=1/(\gamma P_0) \simeq 26.5$ km.

The initial condition chosen for approximating the experimental field reads 
\begin{equation}\label{eq:CI}
  \psi(x,t=0)= \psi_p(x) + \psi_{SG}(x-x_0) \, e^{-i v x/2}
\end{equation}
where   $\psi_p(x)$ describes the short pulse and $\psi_{SG}(x)$ describes the SG that has a group velocity $v= 2 |\delta v_g|^{-1} / (|\beta_2| \gamma P_0)^{1/2} \simeq 4.74$ with respect to the short pulse in the $(x,t)$ plane. 

In the IST theory of the integrable focusing 1D-NLSE (with $\epsilon \equiv 0$), the short pulse (resp. the SG) is characterized by an area (or $L_1$-norm) defined as $\mathcal{A}_P=\int |\psi_p(x)| dx$ (resp. $\mathcal{A}_{SG}=\int |\psi_{SG}(x)| dx$) that determines the number of discrete eigenstates (or solitonic modes) that are embedded in the pulse (resp. the SG) \cite{yang2010nonlinear}. In the experiment $\mathcal{A}_{P}$ and $\mathcal{A}_{SG}$ are measured with good accuracy. However the phase of the complex fields $\psi_{P}(x)$ and $\psi_{SG}(x)$ is not measured, which means that the discrete IST eigenvalues associated with the pulse and the SG cannot be measured. However, the discrete eigenvalues can be estimated using some reasonable assumptions about the analytical expressions for $\psi_{P}(x)$ and $\psi_{SG}(x)$. 

For each of the $29$ experiments realized in a single shot, we compute the complex discrete eigenvalue $\lambda_p=i\eta_p$ (and c.c.) ($i^2=-1$, $\eta_p >0$) associated with the short pulse by assuming that it can be approximated by $\psi_p(x) = a \,  \exp{\left(- x^2/(2w^2) \right)}$, where the parameters $a$ and $w$ are determined using a best fit procedure constrained by the fact that the integral $\int |\psi_{P}(x)| dx$ must be equal to the area $\mathcal{A}_{P}$ measured in the experiment. Applying the same procedure for the SG, we compute an ensemble of $N$ discrete eigenvalues $\lambda_j=i\eta_j- v/4$ (and c.c.) ($j=1,..,N, \ \eta_j >0$) by assuming that the initial box can be approximated by $\psi_{SG}(x) =  b \,  \exp {\left(- x^{2n}/(2L^{2n}) \right)}$ where the real parameters $b$, $x_0$, $L$ and the integer parameter $n$ are determined using a best fit procedure constrained by the fact that the integral $\int |\psi_{SG}(x)| dx$ must be equal to the area $\mathcal{A}_{SG}$ measured in each of the $29$ experiments.

 Fig. 3(a) compares the space shift $\Delta x $ measured in the experiment with the IST  formula \eqref{deltax_IST} which assumes the form 
\begin{equation}\label{eq:shift}
\Delta x = \frac{1}{\eta_p} \sum_{j=1}^N \ln  \left| \frac{i\eta_p+i\eta_j+v/4}{i\eta_p-i\eta_j+ v/4}  \right|
\end{equation}  
on using the well-known expression for the elementary position shift $\Delta(\lambda_p, \lambda_j)= (1/\text{Im} \lambda_p) \ln |\frac{\lambda_p - \lambda_j^*}{\lambda_p - \lambda_j} |$ in the 1D-NLSE two-soliton interaction  \cite{Zakharov:72}. Importantly, the spatial shift arising from the interaction between the tracer soliton and the radiative modes composing the box is neglected in Eq. (\ref{eq:shift}) \cite{Alonso:85a,Alonso:85b,Borghese:18}.

As shown in Fig. 3(a), a good quantitative agreement is obtained between experiment (blue crosses) and Eq. (\ref{eq:shift}) (filled circles). This agreement is slightly improved in numerical simulations of Eq. (\ref{eq:NLSE_adim}) (empty circles) because these simulations incorporate the correction to spatial shift due to the presence of radiative modes in the square boxes. Note that the spatial shift $\Delta x$ is defined in the IST theory as an asymptotic quantity determined at infinite evolution time when the interaction between the tracer soliton and other solitons is negligible. In the experiment, the presence of small dissipation breaks the integrability condition inherent in the IST theory and the spatial shift must be measured at finite evolution time. We have measured the spatial shift at the point where the tracer soliton emerges from the SG, see Supplementary material. Despite this limitation, our results show that the shift described by Eq. (\ref{eq:shift}) is robust to the presence of small perturbative effects. 

\medskip
 Along with the macroscopic spatial shift the nonlinear interaction between the tracer soliton and the SG results in a discernible change in the soliton propagation velocity. From the perspective of the SG kinetic theory \cite{GEl:05,GEl:19,GEl:21}  the effective velocity $s(\lambda_p)$ of a tracer soliton propagating through a SG is given by the {\it equation of state}: 
\begin{equation}\label{eq_state_gen} 
s(\lambda_p)= s_0(\lambda_p)+\int _\Gamma \Delta(\lambda_p, \zeta)f(\zeta)[s(\lambda_p)-s(\zeta)] d \xi d \mu \, .
\end{equation}  
Here $s_0(\lambda_p)=-4 \text{Re} \lambda_p$ is the velocity of the non-interacting (free) tracer soliton and  $f(\zeta)$ is  the {\it density of states} (DOS) of the SG---the distribution function of  solitons with respect to the spectral parameter $\zeta= \xi + i \mu \in \Gamma \subset \mathbb{C}$  \cite{GEl:05,GEl:19,GEl:21,Suret:20,Gelash:19}. The equation of state is a direct consequence of the so-called CRA, a fundamental principle that provides the link between micro- and macroscopic properties of SGs \cite{GEl:21}. It is also at heart of generalized hydrodynamics---the hydrodynamic theory of many-body  quantum and classical integrable systems \cite{Doyon:18, doyon_lecture_2020}, whose intimate connection with kinetic theory of soliton gases has been recently established \cite{Bonnemain:22}. While the validity of the CRA has been mathematically proven for KdV and 1D-NLSE SGs via the asymptotic (thermodynamic) limit of multiphase nonlinear wave solutions \cite{el_thermodynamic_2003,GEl:19}, its  experimental verification is lacking. Such a verification is crucial for the justification of the physical validity of the kinetic theory of soliton gases.

Applied to the configuration of our experiment and written in the reference frame associated with the SG (so that $s_0(\lambda_p)=v$, $s(\zeta) \equiv 0$, $\zeta = i\mu \in \Gamma=[0, ib]$, $\lambda_p=i \eta_p - v/4$) the equation of state \eqref{eq_state_gen} yields for the tracer soliton velocity: 
\begin{equation}\label{eq:kinetic1}
s(\eta_p) = \frac{v}{1-  \frac{1}{\eta_p} \int_0^b  \ln  \left| \frac{i\eta_p+i\mu + v/4}{i\eta_p-i\mu +v/4} \right|  f(\mu) d\mu}
\end{equation}
As discussed in ref. \cite{Gelash:19}, nonlinear wavefields in a dense SG evolving from the box initial data has the DOS given by $f(\mu)=\mu / (\pi \sqrt{b^2-\mu^2})$, $\mu \in [0,b)$.    For such DOS the integral in \eqref{eq:kinetic1} can be evaluated explicitly  to give \cite{GEl:19}  
\begin{equation}\label{trial_vel}
s(\eta_p) = \frac{v \eta_p}{\text{Im} \sqrt{b^2 - \eta_p^2 + \frac{v^2}{16} + i \frac{\eta_p v}{2}}}
\end{equation}

Considering the SG of largest extension ($\mathcal{A}_{SG}=76$, $b=0.634$) interacting with the pulse having spectral parameters $\eta_p=0.617$ and $v=4.74$ (see Fig. 1(c)), Eq. (\ref{trial_vel}) predicts that the velocity of the tracer soliton in the interaction region is $s(\eta_p)=5.26$, which represents a relative velocity change $(s(\eta_p)-v)/v$ of $\sim 11\%$.  This velocity change is illustrated in Fig. 3(b) where the dotted line with the largest slope $1/v\simeq 0.211$ is parallel to the trajectory of the free tracer soliton while the dashed line with the smallest slope $1/s(\eta_p) \simeq 0.19$ is parallel to the trajectory followed by the tracer soliton inside the SG. 

In conclusion, we reported an experiment allowing one to investigate the interaction between a soliton and an optical SG. The observables of this interaction---the macroscopic spatial shift of the tracer soliton and its effective velocity---are favorably compared with the theoretical predictions  of the IST theory and of the kinetic theory of SG. These comparisons provide an important step towards the physical validation of the fundamental  theoretical principles behind the spectral  theory of SGs.

\begin{acknowledgments}
This work has been partially supported  by the Agence Nationale de la
Recherche  through the StormWave (ANR-21-CE30-0009) and SOGOOD (ANR-21-CE30-0061) projects,
the LABEX CEMPI project (ANR-11-LABX-0007),   the Ministry of Higher Education and Research, Hauts de France council and
European Regional Development  Fund (ERDF) through the Nord-Pas de
Calais Regional Research Council and the European Regional Development
Fund (ERDF) through the Contrat de Projets Etat-R\'egion (CPER
Photonics for Society P4S). The authors would like to thank the Isaac Newton Institute for Mathematical Sciences for support and hospitality during the programme ``Dispersive hydrodynamics: mathematics, simulation and experiments, with applications in nonlinear waves''  when part of the work on this paper was undertaken. GE's and GR's work was also supported by EPSRC  Grant Number EP/W032759/1.  GR thanks Simons Foundation for partial support. 

\end{acknowledgments}






%

\pagebreak
\renewcommand{\theequation}{S\arabic{equation}}
\setcounter{equation}{0}
\onecolumngrid

\maketitle

\renewcommand{\theequation}{S\arabic{equation}}
\renewcommand{\thefigure}{S\arabic{figure}}

\begin{center}
{\bf Supplemental material for : \\"Soliton refraction through an optical soliton gas"}\\
\end{center}

\begin{center}
    Pierre Suret,$^1$ Martin Dufour,$^1$ Giacomo Roberti,$^2$ Gennady El,$^2$ Fran\c{c}ois Copie,$^1$ St\'ephane Randoux$^1$
\end{center}

\begin{center}
  {\it $^1$ Univ. Lille, CNRS, UMR 8523 - PhLAM -Physique des Lasers Atomes et Mol\'ecules, F-59 000 Lille, France
  
  $^2$ Department of Mathematics, Physics and Electrical Engineering, Northumbria University, Newcastle upon Tyne, NE1 8ST, United Kingdom
  }
\end{center}

The purpose of this Supplemental Material is to provide some details about the experimental setup and about the experimental methododology.  All equations, figures, reference numbers within this document are prepended with ``S'' to distinguish them from corresponding numbers in the Letter.\\

\tableofcontents

\section{\label{sec1} Detailed description of the experimental setup }

\begin{figure}[h]
    \includegraphics[width=1\textwidth]{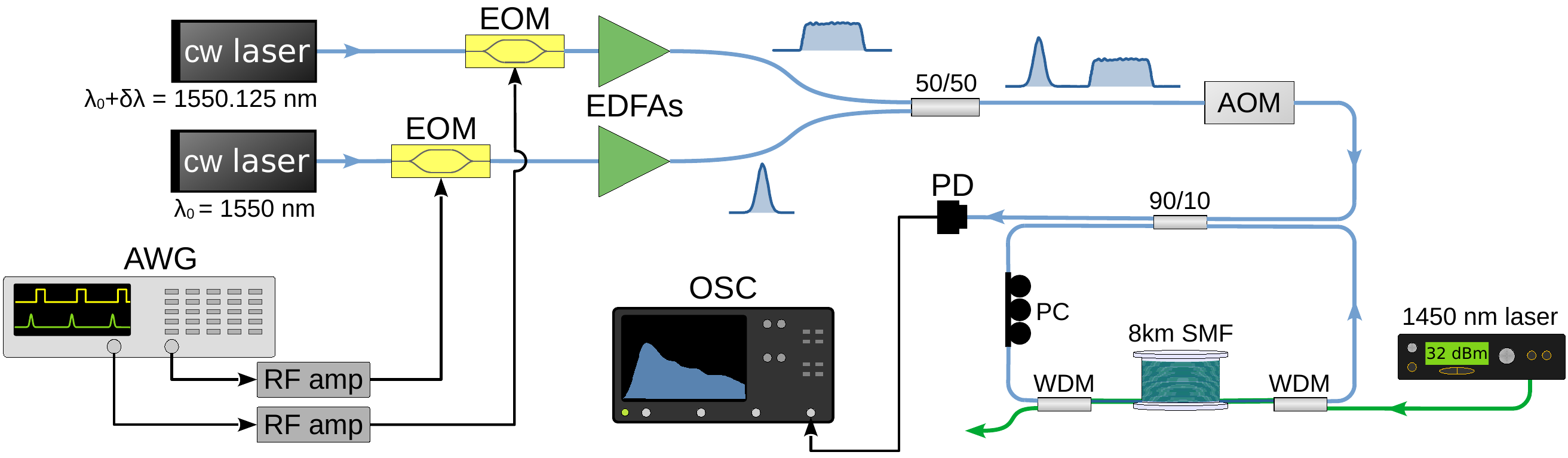}
    \caption{Schematic representation of the experimental setup. The light signals are shaped using two intensity modulators (EOM) driven independently by the two synchronized channels of an arbitrary waveform generator (AWG) having a bandwidth of $25$ GHz. The optical signals are amplified to Watt-level using Erbium-doped fiber amplifiers (EDFAs) before being combined via a $50/50$ fiber coupler and injected inside the recirculating fiber loop. Light detection is made using a fast photodiode (PD) coupled to a fast oscilloscope (OSC) having a bandwidth of $65$ GHz and a sampling frequency of $160$ GSa/s. }
    \label{fig:1}
\end{figure}

Fig. \ref{fig:1} shows a detailed representation of our experimental setup, including some technical details about the generation of the optical signals that are not described in the Letter. The light source used for generation of the short light pulses is a single-frequency continuous-wave (CW) laser diode (APEX-AP3350A) centered at $\lambda_0=1550$ nm which delivers an optical power of a few mW. The short pulses are produced by using a $20$-GHz intensity modulator (iXblue MX-LN-20) connected to an arbitrary waveform generator (AWG) having a bandwidth of $25$ GHz.\\

The optical soliton gases (SGs) are made using another intensity-modulated laser at the wavelength $\lambda_0+\delta \lambda$ with $\delta \lambda=0.125$ nm. The SGs have the initial form of flat-top pulses with durations that increase monotonically from $\sim 200$ ps to $\sim 2000$ ps in $29$ steps. These flat-top pulses are produced using an intensity modulator connected to the second channel of the AWG. This second channel is synchronized with the first channel of the AWG. The initial delay between the short pulses and the flat-top pulses (the optical SGs) can be adjusted with a resolution of $20$ ps by controlling the shapes and delays of the electrical signals delivered by the each of the two channels of the AWG.\\

In our experiment the short pulse and the optical SG have slightly different group velocities because the wavelength of the laser that this used to generate the optical SG is detuned by $\delta \lambda$ from the wavelength of the laser used to generate to short pulses. The group velocity difference $\delta v_g$ between the short pulse and the SG is given by $[\delta v_g]^{-1} \simeq (2\pi c/\lambda_0^2) \beta_2 \, \delta \lambda \simeq -2.16$ ps/km where $c$ represent the velocity of light in vacuum. The group velocity dispersion coefficient of the fiber is $\beta_2=-22$ ps$^2$ km$^{-1}$ at the wavelength $\lambda_0=1550$ nm.\\

The short pulses and flat-top pulses are amplified at the Watt-level by using Erbium-doped fiber amplifiers (EDFAs). Importantly the amplified spontaneous emission (ASE) of the EDFA amplifying the power of the flat-top pulses adds some optical noise, which explains that the generated square pulses evolve into fully randomized SGs inside the recirculating fiber loop. \\

The light signals at the output of the EDFA are combined using a 50/50 fused fiber coupler connected to an acousto-optic modulator (AOM). The AOM plays the role of an optical gate that is open during $\sim 500$ ns for the injection of the light signal into the fiber loop and closed over much longer times (typically $\sim 5$ ms), when the signal circulates inside the fiber loop.  \\

The recirculating fiber loop is made of $\sim 8$ km of single mode fiber (SMF) closed on itself by a $90/10$ fiber coupler. The SMF has been manufactured by Draka-Prysmian. It has a measured second-order dispersion coefficient $\beta_2=-22$ ps$^2$ km$^{-1}$ and an estimated Kerr coefficient $\gamma=1.3$ km$^{-1}$ W$^{-1}$ at the working wavelength of 1550 nm.\\

The $90/10$ coupler is arranged in such a way that $90 \%$ of the intracavity power is recirculated.  The optical signal circulates in the counter-clockwise direction. At each round trip, $10 \%$ of the circulating power is extracted and directed toward a photodetector (PD) coupled to a sampling oscilloscope ($160$ Gsa/s) leading to an overall $32$ GHz detection bandwidth. Experimental data consist in a succession of sequences (one per round trip) that are subsequently processed numerically to construct single-shot space-time diagrams showing the wavefield dynamics.\\

The losses accumulated over one circulation in the fiber loop are partially compensated using a counter-propagating Raman pump at $1450$ nm coupled in and out of the loop via wavelength division multiplexers (WDMs). The pump laser at $1450$ nm is a commercial Raman fiber laser delivering an output beam having a power of several Watt. In our experiments, this optical power is attenuated to typically $\approx$ 200 mW by using a $90/10$ fiber coupler (not shown in Fig. \ref{fig:1}). The mean optical power decay rate of the field circulating in the loop is $\alpha_{eff} \sim 6.2 \times 10^{-4}$ km$^{-1}$ or equivalently $\sim 0.0027$ dB/km. \\

\section{\label{sec2} Circulating optical signal and optical power calibration}

\begin{figure}[h]
    \centering
    \includegraphics{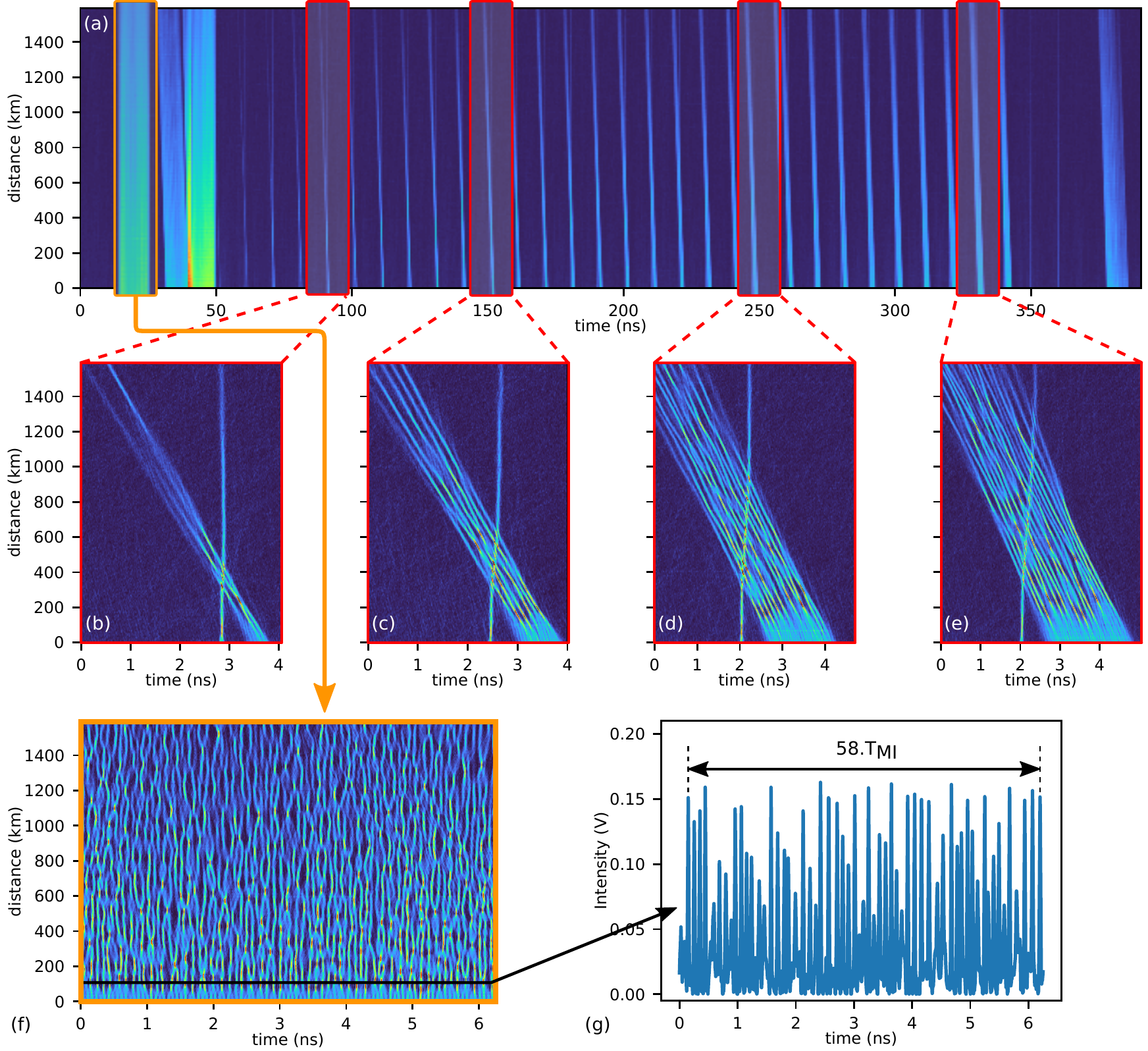}
    \caption{(a) Global space-time pattern recorded at the output of the recirculating fiber loop in a single shot. The region where 29 solitons interact with 29 optical SGs is between $t \simeq 60$ ns and $t \simeq 340$ ns. (b)--(e) Zoomed view on four selected experiments where solitons are refracted by SGs of various extensions. (f) Space-time evolution of a broad pulse of constant power initially pertubated by some optical noise. This space-time pattern is typical from evolution observed in the so-called nonlinear stage of MI. The optical signal measured at $z=96$ km and shown in (f) is used to determine the mean power of the broad pulse. This measurement is used to calibrate the optical power circulating in the fiber loop. }
    \label{fig:2}
\end{figure}

In this Section, we show and describe the characteristics of the entire optical signal that circulates inside the fiber loop. \\

Fig. \ref{fig:2}(a) shows the space-time evolution of the entire signal circulating inside the loop. It has a duration of $\sim 400$ ns and it propagates over a distance of $\sim 1500$ km (the propagation time in one experiment is around $5$ ms). As described in the Letter, the optical signal is composed of an ensemble of $29$ short pulses, each of them being followed by its own SG that has the initial shape of a flat-top pulse perturbated by some optical noise (see Fig. \ref{fig:2}(b)--(e)). \\

In Fig. \ref{fig:2}(a), the time interval devoted to experiments where solitons interact with SGs ranges between $t \simeq 60$ ns and $t \simeq 340$ ns. Between these two times, the duration of the SGs increases monotonically from $\sim 200$ ps to $\sim 2000$ ps (right part of Fig. \ref{fig:2} around $t=330$ ns) in 29 steps. Using this strategy, we capture in one single shot the space-time evolution of a set of $29$ experiments where we observe the interaction between $29$ pulses and $29$ associated SGs of increasing extents. \\

In addition to the broad space-time region where solitons interact with SGs, the optical signal propagating inside the fiber loop has also been designed to incorporate some other regions that permit to measure the optical power circulating inside the fiber loop with a good accuracy. The region in Fig. \ref{fig:2}(a) that is surrounded by an orange rectangle is the region of propagation of a very broad ($\sim$ $10$ ns) flat top pulse perturbated by some small optical noise. As clearly shown in Fig. \ref{fig:2}(f), the flat-top part of the pulse behaves as a plane wave that is destabilized by the small optical noise through the process of modulation instability (MI), as already shown and extensively discussed in ref. \cite{Kraych:19b}.\\

The observed evolution of the nonlinear stage of MI can be used to advantage to measure the optical power circulating inside the fiber loop. Fig. \ref{fig:2}(h) shows the time signal recorded after a propagation distance of $z=96$ km, at a point where large coherent structures can be observed after the initial destabilization of the plane wave. The number of coherent structures observed on a given time span is directly dependent on the period $T_{MI}$ associated with the process of MI. In Fig. \ref{fig:2}(h), we record a total of $58$ coherent structures over a time span of $6.05$ ns, which means that the period associated with the MI process is $T_{MI}= 104.3$ ps. Therefore the mean power $P_0$ of the plane wave in the measurement region is given by $P_0 = (2\pi/T_{MI})^2|\beta_2|/(2\gamma)=  29$ mW \cite{Kraych:19b}. This value is used for calibration and permits to convert the voltage measured by the fast photodiode into an optical power. \\

\begin{figure}[h!]
    \centering
    \includegraphics{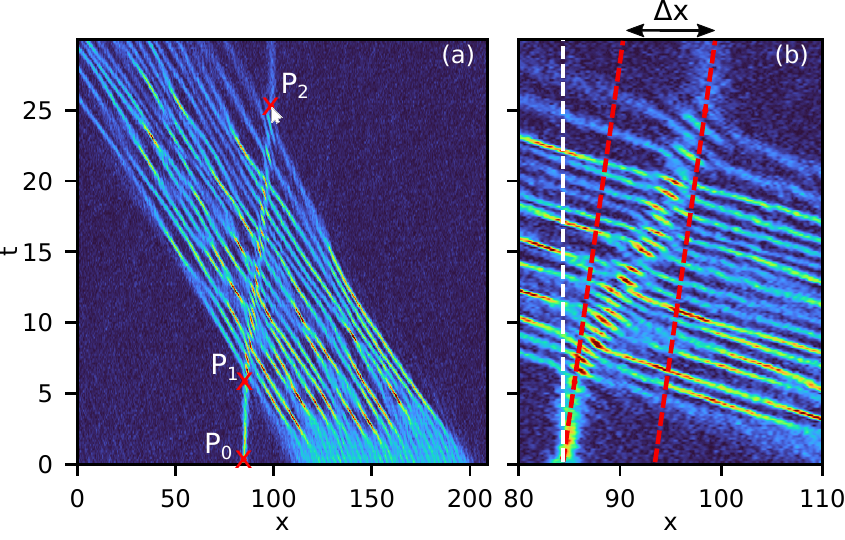}
    \caption{(a) Experimental space-time diagram showing the positions of the points $P_i (x_i,t_i)$ ($i=0,1,2$) that are measured in order to determine the space shift ($\Delta x$) due to the refraction phenomenon. (b) Zoomed view of (a) showing the measured shift $\Delta x$.}
    \label{fig:3}
\end{figure}

\section{\label{sec3}Measurement of the space shift in experiments}

In this section we describe how the space shift associated with the refraction of the soliton by the SGs has been measured in the experiment.\\

In a first step, the space-time diagrams recorded in the experiment are converted to dimensionless units using mathematical transformations given in the Letter, see Eq. (3). In a second step, the position ($x_i,t_i$) of the soliton is measured at three different times ($i=0,1,2)$. The first measurement point is the point $P_0 (x_0,t_0)$ in Fig. 3(a) that corresponds to the smallest time at which the soliton position can be measured. The second measurement point is the point $P_1 (x_1,t_1)$ in Fig. \ref{fig:3}(a) that corresponds to the soliton position just before it is refracted in the SG. The third measurement point is the point $P_2 (x_2,t_2)$ in Fig. 3(a) that corresponds to the soliton position just after it emerges from the SG. \\

The position shift $\Delta x$ computed from simple trigonometric considerations is given by
\begin{equation}
    \Delta x = x_2 -x_0-t_2 \, \tan (\alpha)
\end{equation}
with $\tan (\alpha)=\frac{x_1-x_0}{t_1-t_0}$.

\section{\label{sec4} Experimental measurement of the pulse areas and numerical calculation of the associated discrete IST spectra}

\begin{figure}[h!]
    \centering
    \includegraphics{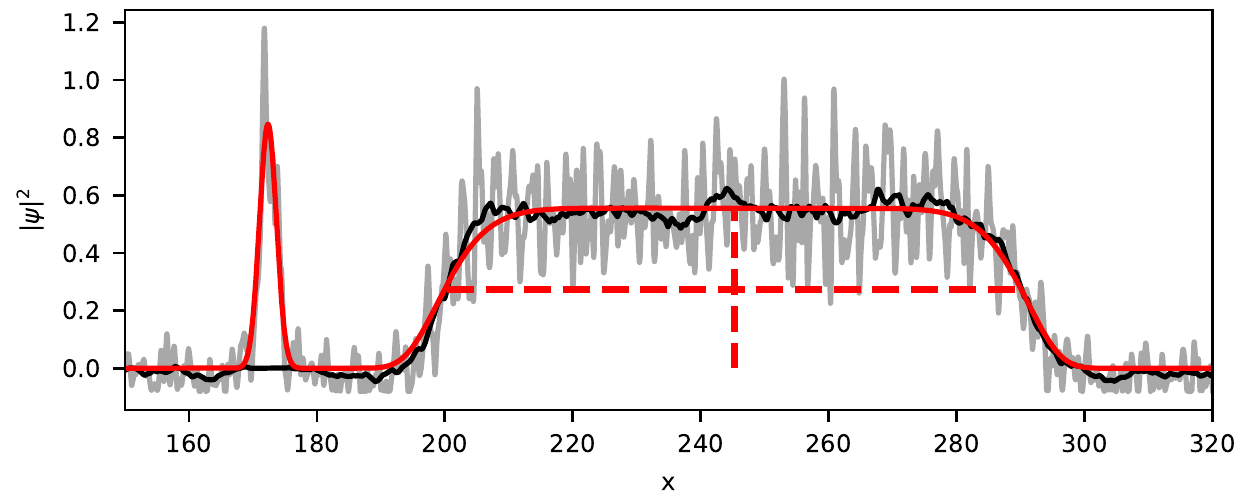}
    \caption{Raw signal recorded in the experiment (gray line) plotted in dimensionless units. The black line represents the raw signal that has been smoothed to remove unwanted oscillations due to the beating between the two laser fields. The red line represents the functions $\psi_p(x)$ and $\psi_{SG}(x)$ that are fitted from the experimental data. }
    \label{fig:4}
\end{figure}

Fig. \ref{fig:4} shows a typical signal that is recorded in the experiment after a propagation over one round-trip ($z=8$ km). In the example shown in Fig. \ref{fig:4}, the signal is plotted in dimensionless units. It consists of one pulse located near one of the SGs of largest extension. The signal plotted with a gray line is the raw signal recorded by the fast photodiode. The fast oscillations at a frequency of $\sim 15$ GHz that are detected on the top of the square pulse are due to the beating between the laser used to produce the short pulses and the laser used to produce the square pulses (the optical SGs).\\

The extinction ratio of the EOMs (see Fig. \ref{fig:1}) is of $20$ dB. This means that residual light carrying a power that is $\sim 1 \%$ smaller than the power of the modulated signals propagate with the square and pulsed signals produced by the modulators. This produces a spectacular beating pattern on the top of the square pulses which represents however an observation artifact. In particular, the observed beating signal has no influence on the solitonic content of the pulses.\\

In order to measure the area of the square pulses, the beating signal is suppressed by using a filter that smooth the unwanted oscillation at $\sim 15$ GHz. This gives the signal plotted with the black line in Fig. \ref{fig:4}.  The area $\mathcal{A}_{SG}$ under the black line defined by $\mathcal{A}_{SG}=\int |\psi(x)| dx$ is then easily computed. As described in the Letter, this signal in black line is then fitted by the function $\psi_{SG}(x) =  b \,  \exp {\left(- x^{2n}/(2L^{2n}) \right)}$ where the real parameters $b$, $x_0$, $L$ and the integer parameter $n$ are determined under the constraint that the integral $\int |\psi_{SG}(x)| dx$ must be equal to the area $\mathcal{A}_{SG}$ that has been measured for the experimental signal.\\

The same procedure is used for the short pulse except that the filtering stage is not applied. The measured profile is fitted by $\psi_p(x) = a \,  \exp{\left(- x^2/(2w^2) \right)}$, where the parameters $a$ and $w$ are determined under the constraint that the integral $\int |\psi_{P}(x)| dx$ must be equal to the area $\mathcal{A}_{P}=\int |\psi(x)| dx$ measured for the short pulse in the experiment. \\

Once the parameters characterizing the experimental pulses (($a,w$) and ($b,L,n$)) are determined, their discrete IST spectra are computed numerically using the Fourier collocation method described in ref. \cite{yang2010nonlinear}.

\end{document}